\documentclass{ws-procs9x6}

\usepackage{amsmath}

\def\al{\alpha}
\def\be{\beta}

\def\de{\delta}
\def\ep{\epsilon}

\def\la{\lambda}

\def\mn{{\mu\nu}}

\def\frac#1#2{{\textstyle{{#1}\over {#2}}}}

\def\lsim{\mathrel{\rlap{\lower4pt\hbox{\hskip1pt$\sim$}}
    \raise1pt\hbox{$<$}}}
\def\gsim{\mathrel{\rlap{\lower4pt\hbox{\hskip1pt$\sim$}}
    \raise1pt\hbox{$>$}}}
\def\sqr#1#2{{\vcenter{\vbox{\hrule height.#2pt
         \hbox{\vrule width.#2pt height#1pt \kern#1pt
         \vrule width.#2pt}
         \hrule height.#2pt}}}}

\newcommand{\beq}{\begin{equation}}
\newcommand{\eeq}{\end{equation}}
\newcommand{\bea}{\begin{eqnarray}}
\newcommand{\eea}{\end{eqnarray}}

\begin{document}

\title{COVARIANT PHOTON QUANTIZATION IN THE SME}

\author{D. Colladay}

\address{New College of Florida,\\
Sarasota, FL 34234, USA\\
$^*$E-mail: colladay@ncf.edu}

\begin{abstract}
The Gupta Bleuler quantization procedure is applied to the SME photon sector.  
A direct application of the method to the massless case fails due to an
unavoidable incompleteness in the polarization states.
A mass term can be included into the photon lagrangian to rescue the
quantization procedure and maintain covariance.
\end{abstract}

\bodymatter

\section{Introduction}
The fermion sector of the SME was quantized consistently during the first stages
of its theoretical development, at least in theories
with a significant nonzero mass parameter \cite{collkos}.
The photon sector has remained largely unaddressed due to several factors that make
it more complicated to deal with.
For example, there is no simple linear Hamiltonian arising from the equation of motion
that can be used for a complete set of orthogonal states.  In addition, the modified equation
of motion has implications for the gauge states that are nontrivial to incorporate.
Addition of a mass term to the lagrangian makes the problem more similar to the fermion
case and generates a tractable problem, so this is the approach used in this talk.
An alternative, perturbative approach in the non-birefringent case has also been implemented
 \cite{hpw}.

\section{Gupta Bleuler method applied to the SME}

The starting point is the Stuckelberg Lagrangian including a CPT-conserving Lorentz-violating term
as well as a mass term
\beq
{\cal{L}} = -{1 \over 4}F_\mn F^{\mn} - {1 \over 4} k_F^{\mu \nu \al \be} F_{\mn} F_{\al \be} 
+{1 \over 2} m^2 A_\mu A^\mu - {\la \over 2}(\partial_\mu A^\mu)^2 .
\label{lagrangian-A}
\eeq
Note that the CPT-violating term has been omitted since it can cause instabilities 
even at tree level \cite{jackiw}.
The gauge condition $\la=1$ is also chosen for simplicity of the commutation relations.
The starting assumptions are the standard covariant commutation relations
for the field and the conjugate momenta
\beq
\pi^j = F^{j0} + k_F^{j0 \alpha \beta}F_{\al \be}, \quad \pi^0 = - \partial_\mu A^\mu .
\eeq
Imposing equal-time canonical commutation rules
\beq
[A_\mu(t,\vec x), \pi^\nu(t,\vec y)] = i \de_\mu^{~\nu} \de^3(\vec x - \vec y),
\eeq
along with
\beq
[A_\mu(t,\vec x), A_\nu(t,\vec y)] = [\pi^\mu(t,\vec x), \pi^\nu(t,\vec y)]=0 ,
\label {comrels1}
\eeq
implements the standard canonical quantization in a covariant manner as is
done in
the conventional Gupta-Blueler method \cite{gb}.
This implies that the time derivatives of the spatial components $A^i$ satisfy
the modified commutation relations
\beq
[\dot A^i(t,\vec x), A^j(t,\vec y)] = - i  R^{ij} \de^3(\vec x - \vec y)
\eeq
where $R^{ij}$ is the inverse matrix of $\de^{ij} - 2 (k_F)^{oioj}$.
In any concordant frame where $k_F$ is reasonably small, this inverse exists.
The commutation relations involving $\dot A^0$ and $A^i$are the same as in the usual
case, so it is convenient to define a covariant-looking
tensor $\overline \eta^\mn$ by setting
$\overline \eta^{00} = 1$, $\overline\eta^{0i} = 0$, and $\overline \eta^{ij} = -R^{ij}$
The commutation relations are expressed as
\beq
[\dot A^\mu(t,\vec x), A^\nu(t,\vec y)] = i \overline\eta^\mn \de^3(\vec x - \vec y) .
\label{comrels2}
\eeq
This matrix is also the inverse of $\tilde \eta^\mn = \eta^\mn -2 k_F^{\mu 0 0 \nu}$
as $\overline\eta^\mn \tilde\eta_{\nu \al} = \eta^\mu_{~\al}$.
Note that the time derivatives of $A$ do not commute, rather
\beq
[\dot A^\mu(t,\vec x), \dot A^\nu(t,\vec y)] = -2 i \overline \eta_{\mu \alpha}
(k_F^{\al 0 i \be} + k_F^{\be 0 i \al})\overline \eta_{\be \nu}{\partial \over \partial x^i}
\de^3(\vec x - \vec y),
\label{comrels3}
\eeq
involving the spatial derivatives of the delta function.

The equation of motion in momentum space is
\beq
(p^2 - m^2) \ep^\mu + 2 (k_F)^{\mu \al \nu \be} p_\al p_\be \ep_\nu = 0,
\eeq
where $\ep^\mu$ is the polarization vector.
One implication of this equation is found by dotting with $p_\mu$ yielding
the condition 
\beq
p^2=m^2 ~~{\rm or}~~ \ep \cdot p = 0 ,
\label{transversality}
\eeq
A key observation is the modified orthogonality
relation for the polarization vectors that follows from the equation of motion
\beq
\ep^{(\la^\prime)}_\mu(\vec p) \left[ \left( p_0^{(\la)} + p_0^{( \la^\prime)}\right) \tilde \eta^\mn 
-2 \left(k_F^{\mu 0 i \nu} + k_F^{\nu 0 i \mu} \right) p_i \right] \ep_\nu^{(\la)}(\vec p)
=2 p_0^{(\la)} \eta^{\la \la^\prime},
\label{orthog1}
\eeq
which holds whenever $p_0^{(\la)} \ne p_0^{(\la^\prime)}$.
The normalization is chosen so that the $\la=0$ polarization vector is timelike while the
others are spacelike.
This is possible due to the presence of a sufficiently large mass term which generically protects the normalization of the polarization vectors from vanishing.
One of the issues of taking the $m \rightarrow 0$ limit is that the above orthogonality condition
can fail due to some polarization vectors becoming light-like.

\section{Momentum-Space Expansion}
The fields can be expanded in a standard Fourier expansion using
\beq
A_\mu(x) = \int {d^3 \vec p \over (2 \pi)^3} \sum_\la {1 \over 2 p_0^{(\la)}}\left( a^\la(\vec p) 
 \ep_\mu^{(\la)}(\vec p)
e^{-i p \cdot x} +  a^{\la\dagger} (\vec p)  \ep_{\mu}^{(\la)}(\vec p)
e^{i p \cdot x}\right).
\label{fouriermodes}
\eeq
The modified orthogonality relation for the polarization vectors can be used
to invert this transform and solve for the raising and lowering operators.  
A straightforward computation then yields the standard relations
\beq
[a^\la(\vec p), a^{\la^\prime \dagger}(\vec q)] = - (2 \pi)^3 2 p_0^{(\la)} \eta^{\la \la^\prime} \de^3(\vec p - \vec q),
\eeq
as well as 
\beq
[a^\la(\vec p), a^{\la^\prime}(\vec q)] = 0,
\eeq
demonstrating that the raising and lowering operators obey conventional statistical relations.
There are subtle issues associated with the above expansion.
Although the mass term is not explicitly present it turns out to be crucial for
generating a complete set of polarization states required for the quantization procedure.
When the mass is set to zero it turns out that the conjugate momentum $\pi^0$ is 
identically zero for most directions in momentum space.  This creates a serious problem for
Gupta-Bleuler as the gauge term initially added into the Lagrangian is not sufficient to 
produce a generically nontrivial conjugate momenta for $A^0$.  This indicates that the
standard Gupta Bleuler method in fact fails in the massless case, 
at least when there is birefringence present.

\section{Explicit Example}
As an explicit example of the issue with $m \rightarrow 0$, consider the single parameter model
$k_F^{0103} = k/2 $ (along with
required nonzero symmetric components to make it anti-self-dual and therefor pure
birefringent).
When $p_1=p_2=0$, the matrix for $K$ in the equation of motion is
\beq
K^{\mn}(p)=
k\left(
\begin{array}{cccc}
 0 &  p_0 p_3 &  0 & 0 \\
  p_0 p_3 & 0 & 0 & -p_0^2 \\
  0 & 0 & 0 & 0 \\
 0 &-p_0^2  & 0 & 0
\end{array}
\right).
\eeq
Searching for zero eigenvalues (with $p^2=0$, candidates for the gauge modes...) yields two eigenvectors, one with the polarization
vector proportional to the momentum and another with 
\beq
\ep = \left(
\begin{array}{c}
0\\0\\1\\0 
\end{array}
\right).
\eeq
Both of these modes satisfy $\ep\cdot p = 0$ indicating that there is in fact no nontrivial mode corresponding to $\pi^0 = - \partial \cdot A $.
Making the momentum more general does not help as the rank of the $K$ matrix is generally
increased to three indicating the same fundamental problem.

\section{Summary}
The standard Gupta-Bleuler method seems to work well when there is a mass
term present in the Lagrangian, but there are serious impediments to implementing
this method when the mass is identically zero.  The most serious issue appears to be
the vanishing of $\pi^0$ implied by the equation of motion, something that is not
an issue in the conventional case.
In addition, certain directions in momentum space yield a set of polarization vectors
that is strictly less than four-dimensional.

\section*{Acknowledgments}
I would like to thank New College of Florida for summer funding that aided in
the completion of this project.

\end{document}